\documentclass[prl,twocolumn]{revtex4}
\usepackage[T1]{fontenc}       

\usepackage{hyperref}
\usepackage{graphicx}
\usepackage{multirow}
\usepackage{color}
\usepackage{amsmath,amsfonts,amssymb}
\newcommand{\nbs }{NbSe$_2$}
\newcommand{\slnbs }{single-layer NbSe$_2$}
\newcommand{\nbsblg }{single-layer NbSe$_2$/BLG}
\newcommand{\etal}{\emph{et al.}~}

\begin{document}

\title{Electronic structure of $2H$-NbSe$_2$ single-layers in the CDW state}

\author{Jos\'{e} \'{A}ngel Silva-Guill\'{e}n}
\email{josilgui@gmail.com}
\affiliation{Fundaci\'on IMDEA Nanociencia, c/Faraday 9, Campus Cantoblanco, 28049 Madrid, Spain}

\author{Pablo Ordej\'on}
\affiliation{Catalan Institute of Nanoscience and Nanotechnology (ICN2), CSIC and The Barcelona Institute of Science and Technology, Campus Bellaterra, 08193 Barcelona, Spain}

\author{Francisco Guinea}
\affiliation{Fundaci\'on IMDEA Nanociencia, c/Faraday 9, Campus Cantoblanco, 28049 Madrid, Spain}
\affiliation{Department of Physics and Astronomy, University of Manchester, Oxford Road, Manchester M13 9PL, UK}

\author{Enric Canadell}
\affiliation{Institut de Ci\`encia de Materials de Barcelona (ICMAB-CSIC), Campus Bellaterra, 08193 Barcelona, Spain}



\begin{abstract}
A density functional theory study of \nbs ~single-layers in the normal non-modulated and the $3\times3$ CDW states is reported. We show that, in the single layer, the CDW barely affects the Fermi surface of the system, thus ruling out a nesting mechanism as the driving force for the modulation. The CDW stabilizes levels lying around 1.5 eV below the Fermi level within the Se-based valence band but having a substantial Nb-Nb bonding character. The absence of interlayer interactions leads to the suppression of the pancake-like portion of the bulk Fermi surface in the single-layer. We perform scanning tunneling microscopy simulations and find that the images noticeably change with the sign and magnitude of the voltage bias. The atomic corrugation of the Se sublayer induced by the modulation plays a primary role in leading to these images, but the electronic reorganization also has an important contribution. The analysis of the variation of these images with the bias voltage does not support a Fermi surface nesting mechanism for the CDW. It is also shown that underlying graphene layers (present in some of the recent experimental work) do not modify the conduction band, but do affect the shape of the valence band of \nbs ~single-layers. The relevance of these results in understanding recent physical measurements for \nbs ~single-layers is discussed.
\end{abstract}

\maketitle

\section{Introduction}\label{introduction}

\begin{figure}
\includegraphics[scale=1]{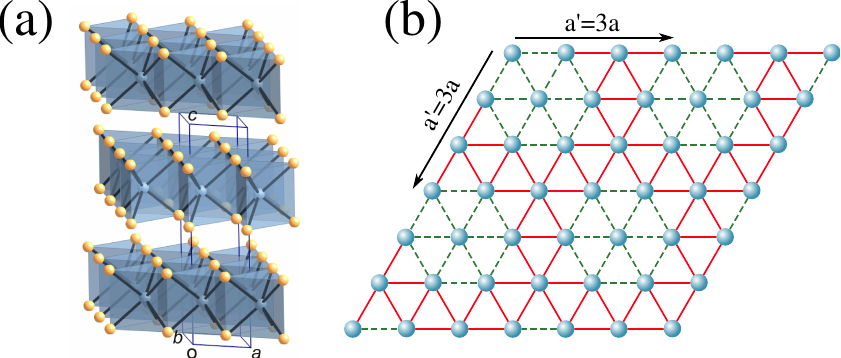}
\caption{(a) Crystal structure of bulk $2H$-NbSe$_2$. (b) Nb layers in the $3\times3$ CDW structure of NbSe$_2$. Nb-Nb contacts shorter/longer than in the average structure are shown as full red/dashed green lines.}\label{fig:CDW_structure}
\end{figure}

Transition metal dichalcogenides are layered materials, easily exfoliable due to the van der Waals forces linking their layers.
They have been the focus of large attention in the past few years because they are ideal systems where to study the influence of the reduced electronic screening brought about by lowering the dimensionality from bulk to layers of different thickness.
Among them, $2H$-\nbs ~(from now on we will refer to it just as NbSe$_2$) is metallic at room temperature, becomes superconducting (SC) at around 7 K~\cite{Revolinsky_1963, Revolinsky_1965} and there are strong indications that it is a two-gap superconductor.~\cite{Yokoya,Boaknin,Huang,Fletcher,Noat2015}
Before reaching the SC state it undergoes a charge density wave (CDW) distortion at around 30 K.~\cite{Wilson,Wilson2}
The bulk structure of \nbs ~is built from hexagonal layers containing Nb atoms in a trigonal prismatic coordination (see Fig. \ref{fig:CDW_structure}a),~\cite{MD2001} but there are also relatively short interlayer Se-Se contacts providing a substantial interlayer coupling.
Although the occurrence of both CDW and SC in this material has been known for a long time,~\cite{Wilson2} the controversy concerning both the nature of the CDW and the competition between CDW and SC states has never been settled. For instance, several structural models of the CDW phase have been discussed but it is only recently that the detailed structure has been reported (see Fig. \ref{fig:CDW_structure}b).~\cite{malliakas_2013} An important question which is the focus of much contemporary debate is how this competition is affected when moving from bulk to single-layer.
It is well established that $T_C$ in \nbs ~diminishes when decreasing the number of layers.~\cite{Frindt, Stanley2009} However, only recently work on single-layers became possible.\cite{Ugeda_2015,Geim_SC_2015} Another layered dichalcogenide where SC and CDW are in competition, although now as a function of doping, is  $1T$-TiSe$_2$. Single layers of this material are also the focus of very much current attention.~\cite{SuNaShi16,ChenChiang2015,Li16}  

Recent experiments \cite{Ugeda_2015,Geim_SC_2015} have shown that $T_C$ lowers down to 1.9 K in \nbs ~single-layers and that the CDW seen in the bulk is preserved. The work by Ugeda \emph{et al.} \cite{Ugeda_2015} using  low-temperature scanning tunnelling microscopy and spectroscopy (STM and STS), transport measurements and angle-resolved photoemission (ARPES) is by now the most comprehensive experimental study of the \nbs ~single-layers.
This work was carried out on high-quality submonolayer \nbs ~films grown on epitaxial bilayer graphene (\nbsblg) on 6H-SiC(0001) and a number of features of the electronic structure were unveiled. An intriguing aspect of this study was the report of a very narrow ($\sim$  4 meV) quasi-gap feature at the Fermi level in the CDW state. Xi and coworkers~\cite{xi_NatNano_2015} reported transport and Raman measurements for layers of different thicknesses and also proposed that the CDW is preserved in the single-layer, although neither the resistivity nor the optical measurements seemed to suggest the opening of any gap at the Fermi level before reaching the superconducting state at $\sim $3 K.

Previous theoretical reports for \nbs ~single-layers either dealt with the undistorted structure of the material~\cite{Lebegue2009} or were based on a CDW structure with a modulation wave vector different from that reported for the bulk,~\cite{Calandra2009} which is at odds with the above mentioned findings.
Thus, appropriate theoretical data to discuss the experimental observations is still lacking.
We have recently considered in detail the electronic structure of bulk \nbs ~and the possible origin of its two-gap superconductivity.~\cite{Noat2015} We were intrigued by some of the experimental results on \slnbs ~and decided to study in detail the electronic structure of the modulated and non modulated structures of \nbs ~single-layers.
In this work, we report first-principles density functional theory (DFT) results which provide a comprehensive description of the electronic structure of this system, bring some light into the long-debated mechanism of the CDW, report in depth calculations of the DOS near the Fermi level to discuss the surprising quasi-band gap feature reported by Ugeda \etal~\cite{Ugeda_2015} and compare our scanning tunneling microscope (STM) simulations with the experimental images, relating them with the structural features of the CDW.
Note that, since the CDW in the bulk is such that there is no definite interlayer ordering, at least some of the conclusions drawn in this work concerning a single-layer should most likely be also valid for the bulk.
In particular, since the STM images are dominated by the top Se layer, we should find similar images for the bulk and $N$-layer systems.
We also consider whether the inclusion of a graphene supporting layer in the calculations could affect some of these results.

\begin{figure}
  \includegraphics[scale=0.8]{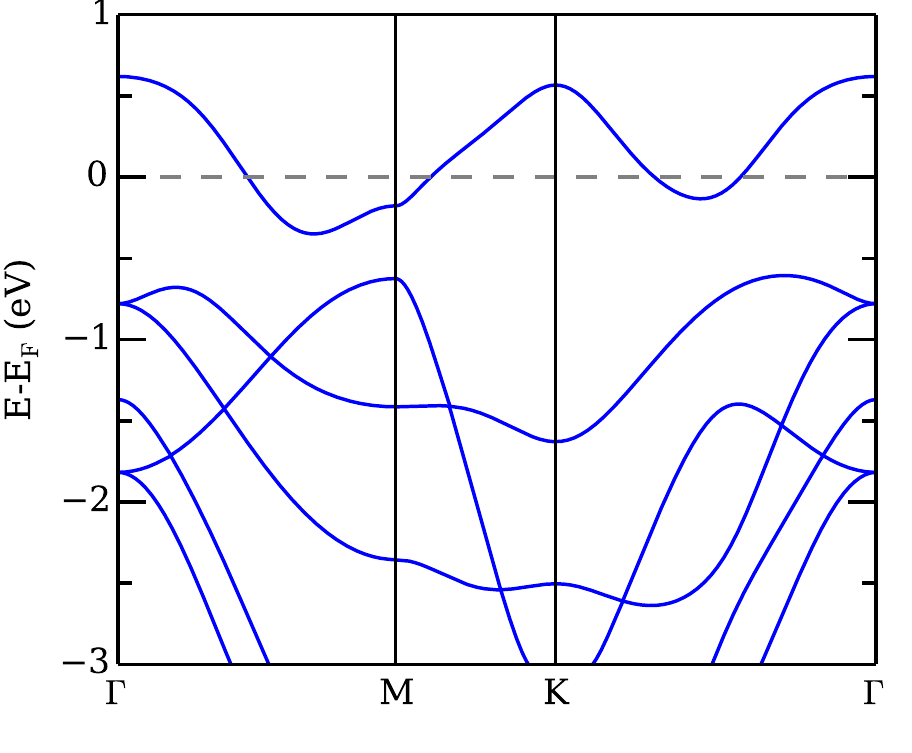}
\caption{Band structure for non-modulated \slnbs, obtained using the average structure for the bulk.~\cite{malliakas_2013}}\label{fig:BS_1x1}
\end{figure}

\begin{figure*}
\includegraphics[scale=0.9]{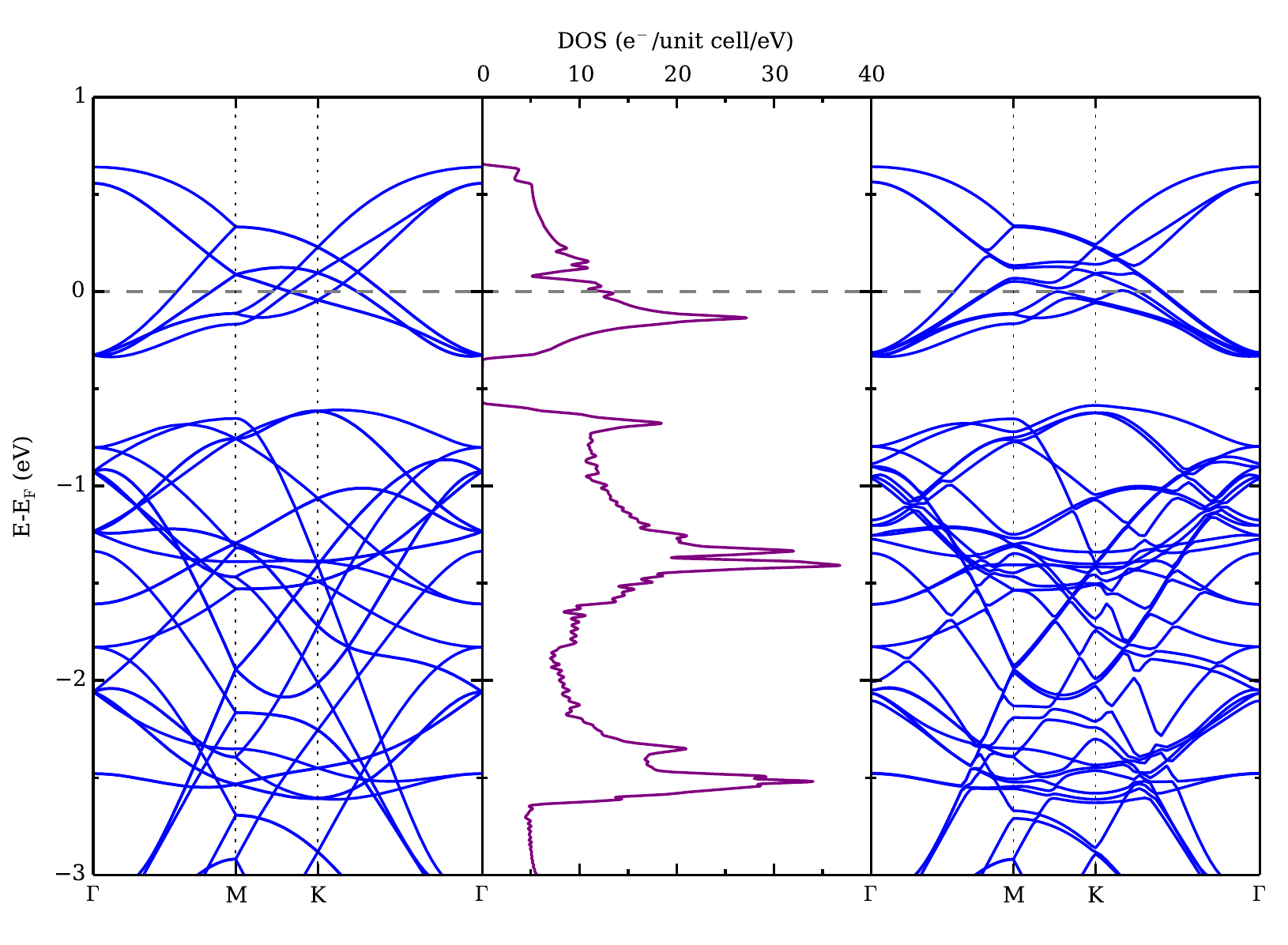}
\caption{Band structure for non-modulated (left) and $3\times3$ modulated (right) \slnbs. DOS for the modulated structure (center). The calculations are made on the basis of the average and modulated structures for the bulk.~\cite{malliakas_2013}}\label{fig:BS_DOS}
\end{figure*}

\section{Results and Discussion}

As a first step in our study, we search for structures with $3\times3$ periodicity by performing structural relaxations of the atomic coordinates of a $3\times3$ supercell of \slnbs. These calculations, using the experimental in-plane lattice constant of the bulk modulated structure, led to the same distortion pattern as in Fig. \ref{fig:CDW_structure}b. However, we have also located some structurally related local minima with the same symmetry, but with slightly different distortion patterns. These structures are associated with energies lying within a very narrow energy range of 2 meV $per$ \nbs ~formula unit, thus suggesting a very flat potential energy surface, consistent with the presence of the CDW distortion at low temperature. We have also found that the DOS calculated for these minima do not exhibit significant differences. Consequently, all calculations reported here for the CDW state are based on the experimental structure reported for the CDW in the bulk.~\cite{malliakas_2013}

Fig.~\ref{fig:BS_1x1} shows the band structure of the normal phase of the \nbs ~single layer. The bulk phase has two layers per unit cell, so each of the bands present in the single layer splits in two for the bulk (see, for instance, Fig. 8 in Ref. \citenum{Noat2015}).
A clear difference with the bulk is that, in the single-layer, a gap opens from around 0.35 to 0.6 eV below the Fermi level,  between the conduction band crossing the Fermi level (based on Nb $d$-orbitals) and the lower-lying valence bands. Due to the lack of (mainly Se-Se) interlayer interactions, the highest Se-based valence band is noticeably lower in energy in the case of the single layer. In the bulk case, these interactions raise the antibonding Se band, which then closes the gap and overlaps with the Nb conduction bands, even crossing the Fermi level and leading to a small pocket of pancake shape in the Fermi surface.

Each of the bands for the undistorted $1\times1$ phase is folded into a set of nine bands in the $3\times3$ periodicity. Therefore, for an easier comparison between the normal and modulated structures, we show in Fig.~\ref{fig:BS_DOS}a and b the band structure of the undistorted and the CDW phases, respectively, both represented in the Brillouin zone of the $3\times3$ periodicity.  
The presence of the $3\times3$ CDW modulation induces a minor distortion of the band structure of \slnbs, mainly the opening of small gaps and band splittings.

\begin{figure*}
\includegraphics[scale=0.9]{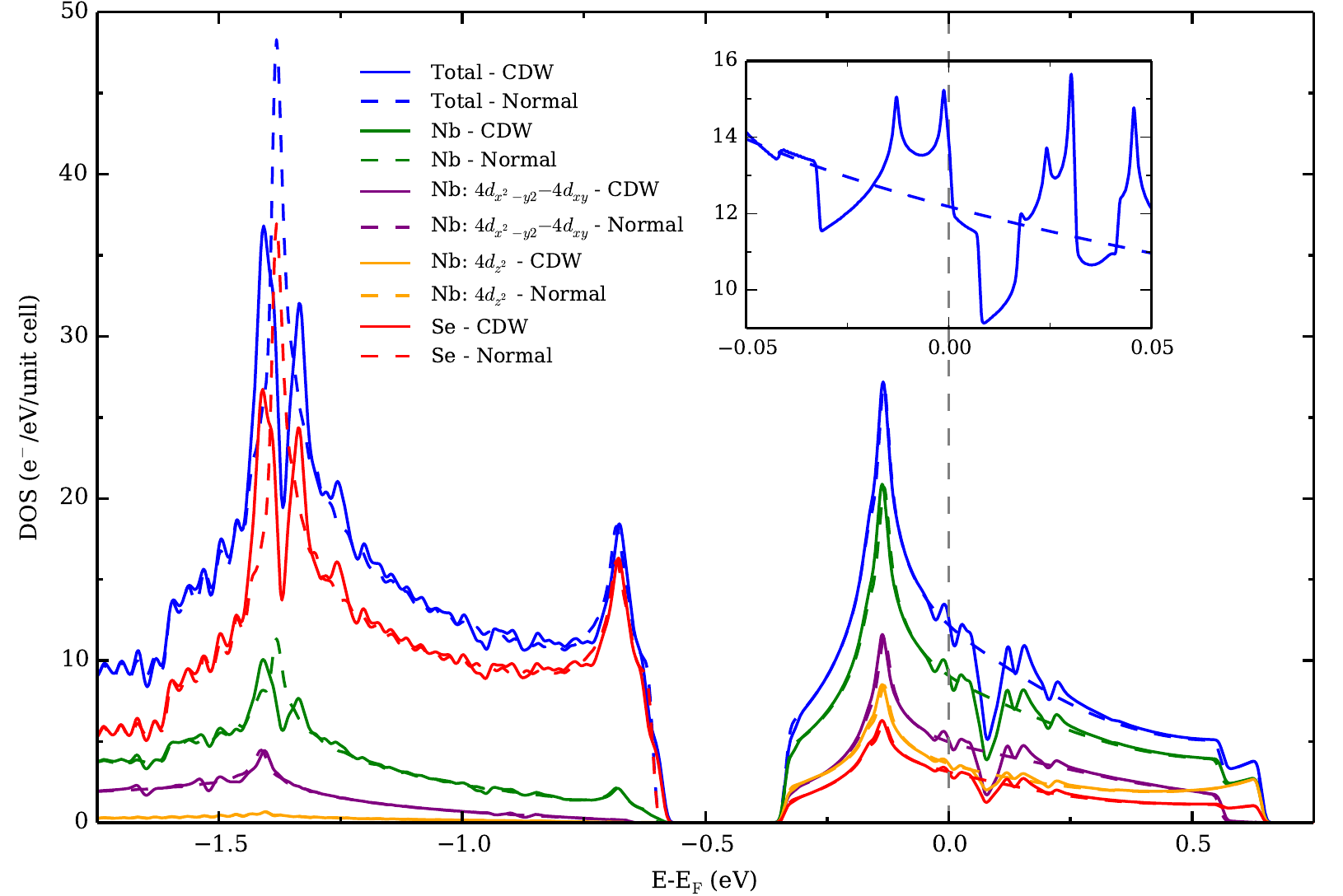}
\caption{Comparison of the DOS and the different Nb and Se local projections for an \nbs ~single-layer with the CDW (full lines) and normal non-modulated (dashed lines) structures. Shown in the inset is DOS in a narrow energy range around the Fermi level calculated with a grid of $1000\times1000\times1$ $k$-points.}\label{fig:DOS_Exp}
\end{figure*}

In Fig.~\ref{fig:DOS_Exp} we show the comparison of the DOS (and the partial contributions from the Nb and Se orbitals) of the normal and modulated structures, computed using a grid of $100\times100$ k-points to sample the Brillouin zone of the $3\times3$ supercell.
Note that the partial DOS associated with both Nb and Se have roughly the same overall shape as the total DOS.
The contribution of the Se orbitals to the DOS of the Nb-based conduction band of Fig. \ref{fig:BS_DOS} is of $\sim 25\%$. Since the orbital character of this band changes around the Brillouin zone (the $z^2$ character prevails around $\Gamma $  whereas the $x^2$-$y^2$/$xy$ character dominates around $K$)~\cite{JMH06,Noat2015} we have also plotted the two separated Nb contributions in this Figure.
The total DOS associated with the Nb band has a strong maximum around 0.12 eV below the Fermi level and then, except for some structure around the Fermi level and up to 0.2 eV, the general shape of the DOS is that of a slowly decreasing function.
For the bands just below the energy gap mentioned before, the states are mainly Se-based, but with a non-negligible contribution from Nb orbitals. 

As observed already with the band structure, the general shape of the DOS for the normal and modulated   structures is similar, although some clear changes occur from -0.05 to +0.25 eV and from -1.3 to -1.4 eV. The changes at the Fermi level are, however, relatively small. A zoom of the region around the Fermi level is shown in the inset of Fig.~\ref{fig:DOS_Exp}, computed using a much finer grid of $1000\times1000$ k-points and narrower gaussian smearing, to better resolve the fine features at this region. The results show changes in the DOS, related to the opening of small gaps in the band structure in the range of energies close to the Fermi level. However, we do not obtain a clear, narrow gap or quasi-gap at $E_F$, as observed experimentally in the study of Ugeda \etal~\cite{Ugeda_2015}. They obtain a very narrow ($\sim $4 meV) dip in their STS spectra at zero voltage at temperatures below the CDW transition, which they interpret as the opening of a gap associated to the CDW distortion. Our results, therefore, do not support this interpretation, at least in the  context of one-electron electronic structure theory. The dip in the DOS observed in our calculations about 10 meV above $E_F$ is too broad compared to the one observed in the experimental STS images, it is not located precisely at $E_F$, and it is not robust with respect to changes in the structural distortion (the optimized structures mentioned before do not reproduce this dip at the same energy, although they do present others centered at different energies and with different widths).  

We now focus on the origin of the general shape of the DOS in the region of the Nb-based conduction band, as well as the noted variations due to the CDW. This will provide some clues concerning the origin of the CDW. 
Except for the structure between -0.05 and 0.25 eV, the general shape of the DOS for the undistorted and CDW structures is the same.
This shape is easy to understand by considering the band structure of Fig.~\ref{fig:BS_1x1}.
From the onset shoulder of the DOS at the minimum of the band in the $\Gamma \rightarrow M$ line, the DOS increases strongly and continuously up to the strong peak around -0.12 eV due to the decreasing slope of the band around $M$, and to the presence of the minimum along the $K \rightarrow  \Gamma $ line.
The DOS then goes down continuously until the top of the band, where a double shoulder is visible due to the two maxima at $\Gamma$ and $K$.  
Despite the small slope  of the bands around these regions, they only lead to a moderate value of the DOS because of the comparative small weight of these points in the Brillouin zone.

The unusual shape of the partially filled Nb-based band, with clear minima in between special points (which ultimately controls the DOS shape), is due to the fact that the Nb $d$ character ($z^2$ and $x^2$-$y^2$/$xy$) changes along the Brillouin zone.~\cite{JMH06,Noat2015} The $x^2$-$y^2$/$xy$ in-plane orbitals are well directed to interact with the same orbitals of neighboring Nb atoms; in contrast, the $z^2$ orbital has lobes pointing outside the layer plane.
Consequently, the former can generate metal-metal interactions but not the latter.
By symmetry, the two types of orbitals can not mix at $\Gamma $ but within the Brillouin zone the two sets can smoothly mix and interchange character between different regions.~\cite{WC1992} Thus, the phase changes associated with the crystal orbitals of some regions of the Brillouin zone lead to bonding interactions between Nb atoms (the regions around the minima) whereas in the region around $\Gamma $ where the $z^2$ character prevails, such bonding interactions are not possible and the energy is higher.

As shown in Fig.~\ref{fig:DOS_Exp}, the most prominent change in the DOS above the Fermi level brought about by the CDW is the occurrence of a noticeable minimum $\sim $ 0.08 eV above the Fermi level.
The comparison between the right and center panels of Fig.~\ref{fig:BS_DOS} shows that the origin of this minimum lies in the region of the slightly avoided crossings occurring not far from the $K$ point along the $\Gamma \rightarrow K$ and $K \rightarrow M$ lines, and around the $M$ point in the reduced Brillouin zone.
However, other bands are dispersive in the energy range of these avoided crossings and no true gap opens.
As expected, a fat-band analysis of the band structure showed that this region is mostly associated with the $x^2$-$y^2$/$xy$ orbitals and consequently, with the metal-metal interactions along the layer.
Since these states are unoccupied they can not provide a stabilization to the system.
However, a counterpart of the electronic rearrangement associated with this minimum is found in the block of bands of the valence band.
Looking at the region below the gap around -0.5 eV from $E_F$ in Fig. \ref{fig:BS_DOS} it is clear that the most important difference affects the large DOS peak around -1.35 eV for the non-modulated structure, which is split into two contributions in the CDW structure.
This peak is associated with a series of bands that, although being Se-based, have an important participation of the Nb $x^2$-$y^2$/$xy$ orbitals making Nb-Nb bonding interactions (and, although not shown in Fig. \ref{fig:DOS_Exp}, a sizeable Nb $p_x,p_y$ contribution).
When this peak splits in the CDW structure, the lower peak is mostly associated with the Nb-Nb bonds which have been shortened in the CDW structure (see red lines in Fig. \ref{fig:CDW_structure}) and thus provide a stabilization to the layer, whereas the upper peak is mostly associated with the Nb-Nb bonds which have been lengthened in the CDW structure (see dashed green lines in Fig. \ref{fig:CDW_structure}) and consequently provide a destabilization.
Since the stabilized levels have a stronger weight in the DOS than the destabilized levels, the distortion associated to the CDW provides a small but definite stabilization to the system.
We thus conclude that the CDW modulation in \nbs ~single-layers is unrelated to the Fermi surface and stabilizes levels which, even being in the Se block, have a substantial extended Nb-Nb bonding character. 
Alternatively, we can describe the tendency towards a CDW state by calculating, in second order perturbation theory, the gain in electronic energy induced by a small displacement towards the CDW state. The previous discussion implies that this calculation is dominated by the CDW induced transitions between the filled states at $E \approx - 1.4$ eV, at the $M$ point in the bands shown in Fig. \ref{fig:BS_DOS}, and the empty states at $E \approx 0.1$ eV. As mentioned above, the occupied states are mostly Se orbitals, while the empty states reside mostly on Nb orbitals. The displacements associated to the CDW bring closer Se and Nb atoms, so that it can be expected that transitions between orbitals in the two atoms are induced.

The easiness of this low temperature metal-metal bonding modulation should decrease with the replacement of selenium by sulfur (less polarizable and smaller), but should increase with the replacement of neobium by tantalum, because of the more extended $d$ metal orbitals. This conclusion is consistent with the bulk onset temperatures for the CDW of different $2H$-MX$_2$ systems (32 K for NbSe$_2$, 122 K for TaSe$_2$ and 80 K for TaS$_2$), whereas there is no CDW condensation for $2H$-NbS$_2$.~\cite{DiSalvo_1977}

\begin{figure}
  \centering
  \includegraphics[scale=0.5]{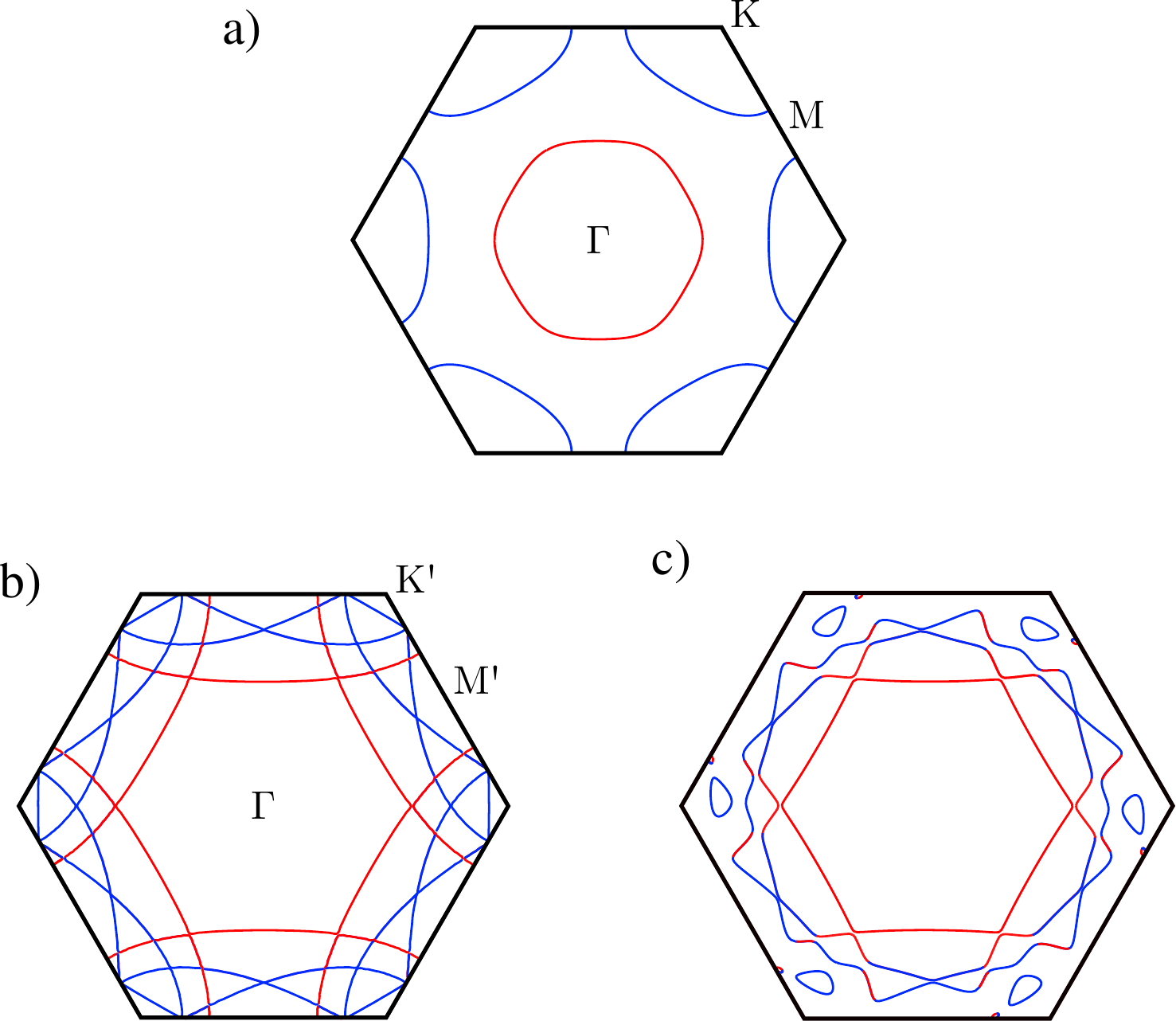}
  \caption{Fermi surface calculated for a \nbs ~single-layer using the non modulated structure (a) and (b) and the $3\times3$ CDW structure (c). The Brillouin zone used in (a) is that appropriate for the undistorted structure whereas that in (b) and (c) is that appropriate for the $3\times3$ CDW structure. Consequently the Brillouin zone in (a) should be nine times larger than actually drawn.}
  \label{fig:FS_Comp}
\end{figure}

\begin{figure*}
\includegraphics[scale=0.9]{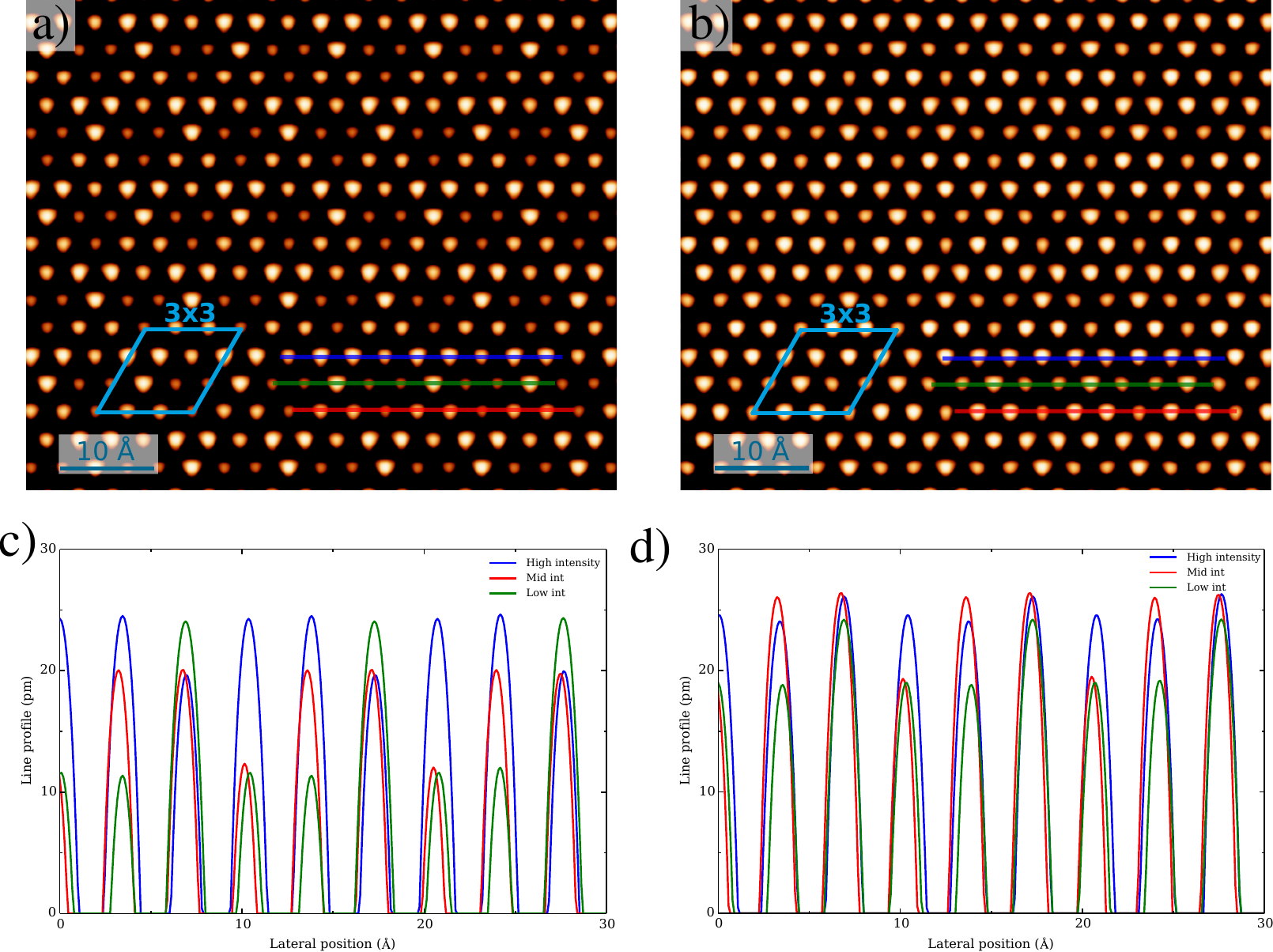}
\caption{STM images of the  $3\times3$ modulated single layer \nbs ~calculated for positive ($V_s$ = +50 mV) (a) and negative ($V_s$ = -50 mV) (b) bias voltages and an iso-DOS value of 9$\times$10$^{-8}$ (e$^-$/eV/unit cell). The profiles along the three directions shown in red, green and blue in (a) and(b) are shown in (c) and (d), respectively.}\label{fig:STM_mod_bias}
\end{figure*}

The calculated Fermi surfaces for the \nbs ~single-layer before and after the $3\times3$ CDW are reported in Figs. \ref{fig:FS_Comp}a and \ref{fig:FS_Comp}c. To facilitate the comparison, the Fermi surface of the undistorted phase is redrawn using the same color code and the nine times smaller Brillouin zone appropriate for the distorted phase in Fig. \ref{fig:FS_Comp}b. The Fermi surface of Fig. \ref{fig:FS_Comp}a is similar to that of the bulk with the notable exception of the absence of the closed, pancake shape portion centered at $\Gamma$ present in the bulk material~\cite{JMH06,Noat2015,{CoMeOnu94}}. This Fermi surface pocket is thus a purely three-dimensional feature.
As a consequence, the area of the two closed portions in Fig. \ref{fig:FS_Comp}a differs from the average cross section of the corresponding cylinders in the bulk Fermi surface. Comparing the Fermi surfaces of Figs. \ref{fig:FS_Comp}b and \ref{fig:FS_Comp}c it is clear that the effect of the CDW on the Fermi surface of the system is really minor, creating very small gaps where the folded lines cross. Thus, as it occurs for the solid and in agreement with the resistivity measurements for the single layer, \cite{Ugeda_2015,Geim_SC_2015,xi_NatNano_2015} the occurrence of the CDW should be practically imperceptible in the transport measurements. These results are thus consistent with our analysis above of the DOS in ruling out the Fermi surface nesting as the possible origin of the CDW. Note in addition, that since the areas of the closed portions/cylinders in the Fermi surface of the single-layer and the bulk differ because of the occurrence of the pancake for the bulk, it would be very unlikely that both the bulk and single-layer phases could exhibit the same commensurate modulation if nesting was at its origin.
We note also that, according to a recent proposal for bulk $2H$-NbSe$_2$,~\cite{Noat2015} a strong quasi-particle coupling should occur between the Se-based pancake and the K-centered cylinders of the Fermi surface leading to the larger of the two superconducting gaps. The absence of the Se-based pancake component of the Fermi surface for single-layer \nbs ~must be most likely related to the variation of its superconducting properties.

The fact that the $3\times3$ CDW order in the bulk remains intact for the single layer of \nbs ~has been made clear through the comparison of    the bulk and single-layer STM topographic images at temperatures below the transition.~\cite{Ugeda_2015} Here we discuss the nature of these STM  images and the relationship with the structural modifications induced by the $3\times3$ modulation. Our images, computed using the Tersoff-Hamman approximation,~\cite{TerHam85} correspond to constant current images, showing the maps of heights that produce a constant tip-surface current. The STM images for single-layer \nbs ~in the CDW phase calculated for positive and negative bias voltages ($\pm$ 50 mV) are shown in Figs.~\ref{fig:STM_mod_bias}a and~\ref{fig:STM_mod_bias}b, respectively. The image for the non-modulated structure is an hexagonal array of identical spots associated with the sublayer of Se atoms at the surface. The calculated images for the $3\times3$ modulated structure with either positive  or negative bias voltages are markedly different, clearly showing the new periodicity. They are consistent with the experimental images for the   bulk~\cite{DaCaAll14} and single-layer \nbs~\cite{Ugeda_2015} modulated structures. In particular, the images change with the sign of the bias    voltage as reported for the bulk.~\cite{DaCaAll14} The pattern exhibited by these images, as well as the bulk~\cite{DaCaAll14} and single-layer~\cite{Ugeda_2015} experimental ones, is roughly the same: a series of triangular units of spots along lines parallel to the diagonal         directions of the hexagonal lattice with essentially three different intensities.

\begin{figure}
\includegraphics[scale=0.8]{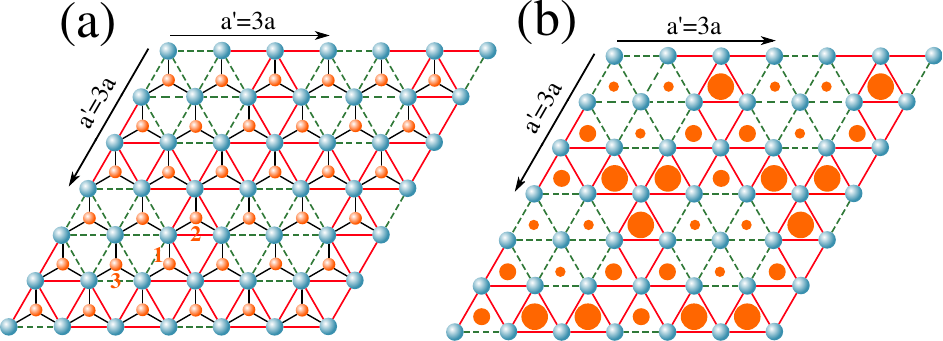}
\caption{(a) $3\times3$ CDW structure of \slnbs. Blue (orange) balls represent Nb (Se) atoms. Nb-Nb contacts shorter/longer than in the average structure are shown as full red/dashed green lines. (b) Schematic STM image predicted for the $3\times3$ CDW structure taking only into account the height of the different Se atoms.}\label{fig:Scheme_STM}
\end{figure}

In principle, this pattern can be thought as reflecting the different heights of the Se atoms induced by the modulation. In the $3\times3$ CDW  structure there are three different types of Se atoms with different heights (see Se$_1$, Se$_2$ and Se$_3$ in Fig. \ref{fig:Scheme_STM}a).
Consideration of the bulk modulated structure as well as our geometry optimized single-layer modulated structure shows that, as intuitively guessed, the height of the Se atoms which lie at the center of a triangle of Nb-Nb distances shorter/longer than in the non-modulated structure (continuous red/dashed green lines in Fig. \ref{fig:Scheme_STM}) is larger/smaller.
Thus, Se atoms 2 and 3 are the highest and lowest lying, respectively.
The height of  Se atoms 1, which are at the center of a triangle with two shortened and one elongated Nb-Nb distances, is in between.
A schematic picture of the STM topographic image, were the intensity of the spots associated with Se$_1$, Se$_2$ and Se$_3$ is represented by large, medium and small circles, respectively, is shown in Fig. \ref{fig:Scheme_STM}b.  This qualitative picture exhibits the same pattern occurring in the experimental images,   with three successive triangular units of spots with different intensities along lines parallel to the lattice diagonal directions.

Looking at the images of Figs. \ref{fig:STM_mod_bias}a and \ref{fig:STM_mod_bias}b, it can be seen that this pattern is indeed observed in our simulated images for both bias voltages. The Se sublayer corrugation induced by the CDW does indeed seem to play a primary role in leading to intensity patterns of the STM images: the more/less brilliant spots are associated with the Se atoms sitting on top of a triangle of Nb atoms with the shorter (Se$_2$)/longer (Se$_1$) Nb-Nb contacts induced by the modulation.
However, looking more carefully at the profiles it becomes clear that the pattern of heights of the images changes with the bias voltage.
This points out that the electronic rearrangement induced by the modulation also plays a non-negligible role.
The profiles along the red, green and blue lines in Fig. \ref{fig:STM_mod_bias}a provide a detailed description of the features of the topographic image.
The ratio between the two maxima along the blue profile is a measure of the intensity ratio of the spots due to the Se$_2$ and Se$_1$ atoms.
In the same way, the ratio between the two maxima along the green profile is a measure of the intensity ratio of the spots due to Se$_2$ and Se$_3$ atoms.
When moving from positive to more negative bias voltages the images evolve in such a way that the difference between the two maxima along the blue profiles decreases (in fact, for -50 mV the ratio is actually slightly reversed, see Fig. \ref{fig:STM_mod_bias}d).
At the same time, the ratio between the two maxima along the green profile decreases (for -50 mV  the ratio has been reduced by approximately one-half with  respect to that for + 50 mV, see Fig. \ref{fig:STM_mod_bias}d). 
Consequently, these calculations confirm that, although the pattern of the STM images is primarily governed by the corrugation of the Se sublattice, the electronic rearrangement induced by the modulation plays also a relevant role, the effect being more visible for the  selenium  atom Se$_1$.

In fact, these calculations suggest that, for some bias voltages, the image pattern can only partially reflect the corrugation of the Se sublayer.
Note, for instance, that especially for positive bias voltages, where the empty states are probed, the region with large differences in the DOS induced by the structural modulation is explored.
Consequently, differences in the STM images as a function of the bias voltage would  not be surprising, although our study suggests that the essential features due to the structural pattern should remain.
It is important to point out that, in agreement with our  analysis of the DOS and Fermi surface, the difference in the STM images found for opposite bias voltages does not support a Fermi surface nesting based mechanism for the $3\times3$ modulation. Although the electronic modulation is always accompanied by a periodic lattice distortion, if    the electronic modulation is dominant one expects complementary images for positive/negative voltage biases, as found for instance in the case of NbSe$_3$~\cite{BruWaMon} or the potassium purple bronze,~\cite{MaZiChev} two typical Fermi surface nesting driven CDW materials. This is clearly  not the case here. Our study thus indirectly confirms the bulk experimental results by Dai \emph{et al.} ~\cite{DaCaAll14}

\begin{figure*}
\includegraphics[scale=0.8]{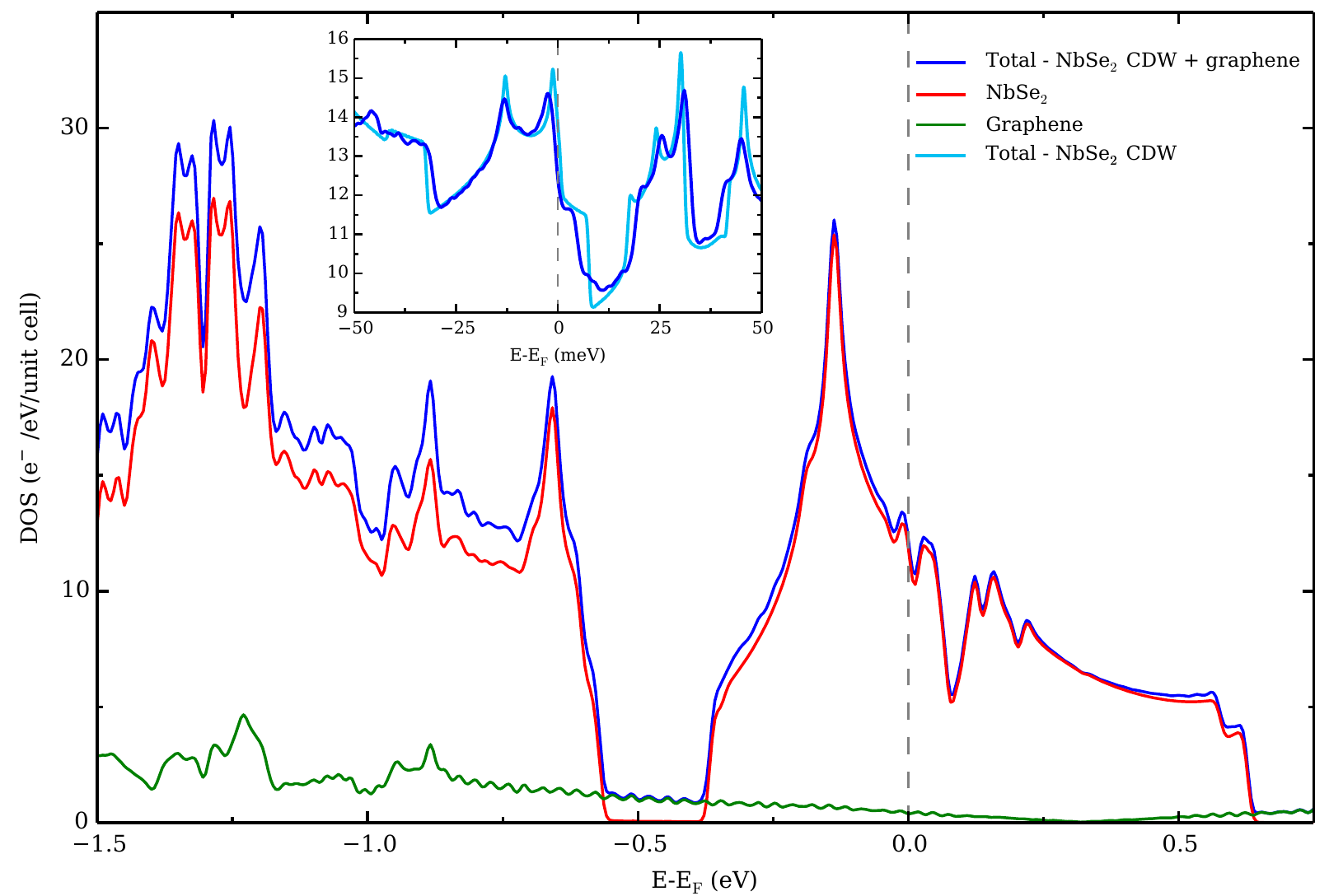}
\caption{Calculated DOS and local projections for the graphene/single-layer \nbs ~system.  Shown in the inset is the comparison of the system with and without graphene DOS in a narrow energy range around the Fermi level calculated with a grid of $500\times500\times1$ $k$-points ($1000\times1000\times1$ in the case without graphene).} \label{fig:NbSe2_graphene}
\end{figure*}

As the recent experimental work by Ugeda \emph{et al.} \cite{Ugeda_2015} was carried out on single-layer \nbs ~supported on bilayer graphene on SiC, we now consider whether the underlying graphene can induce changes on the electronic structure of pristine \slnbs. We carried out calculations for \nbs ~with an underlying graphene layer at an interplanar distance optimized for the non-modulated structure (see Computational Details). The calculated DOS for single-layer \nbs ~on graphene is shown in Fig. \ref{fig:NbSe2_graphene}. The region of the conduction band, including that around the Fermi level (see inset in Fig. \ref{fig:NbSe2_graphene}), is practically unaltered by the presence of graphene. Except for small shifts in the lower part of the valence band, the most important change in this region is the appearance of an additional peak at $\sim $ -0.9 eV. This additional peak results from the interaction of Se $p$ based levels of the \nbs ~single-layer with $\pi $ type states of graphene. We also note a very weak charge transfer from graphene to the \nbs ~single-layer. As far as the electronic structure of the \nbs ~single-layer is concerned, we thus can conclude that the underlying graphene barely affects the region of the Nb-based conduction band but introduces additional peaks and small shifts in the Se-based region of the DOS.

\begin{figure*}
\includegraphics[scale=0.7]{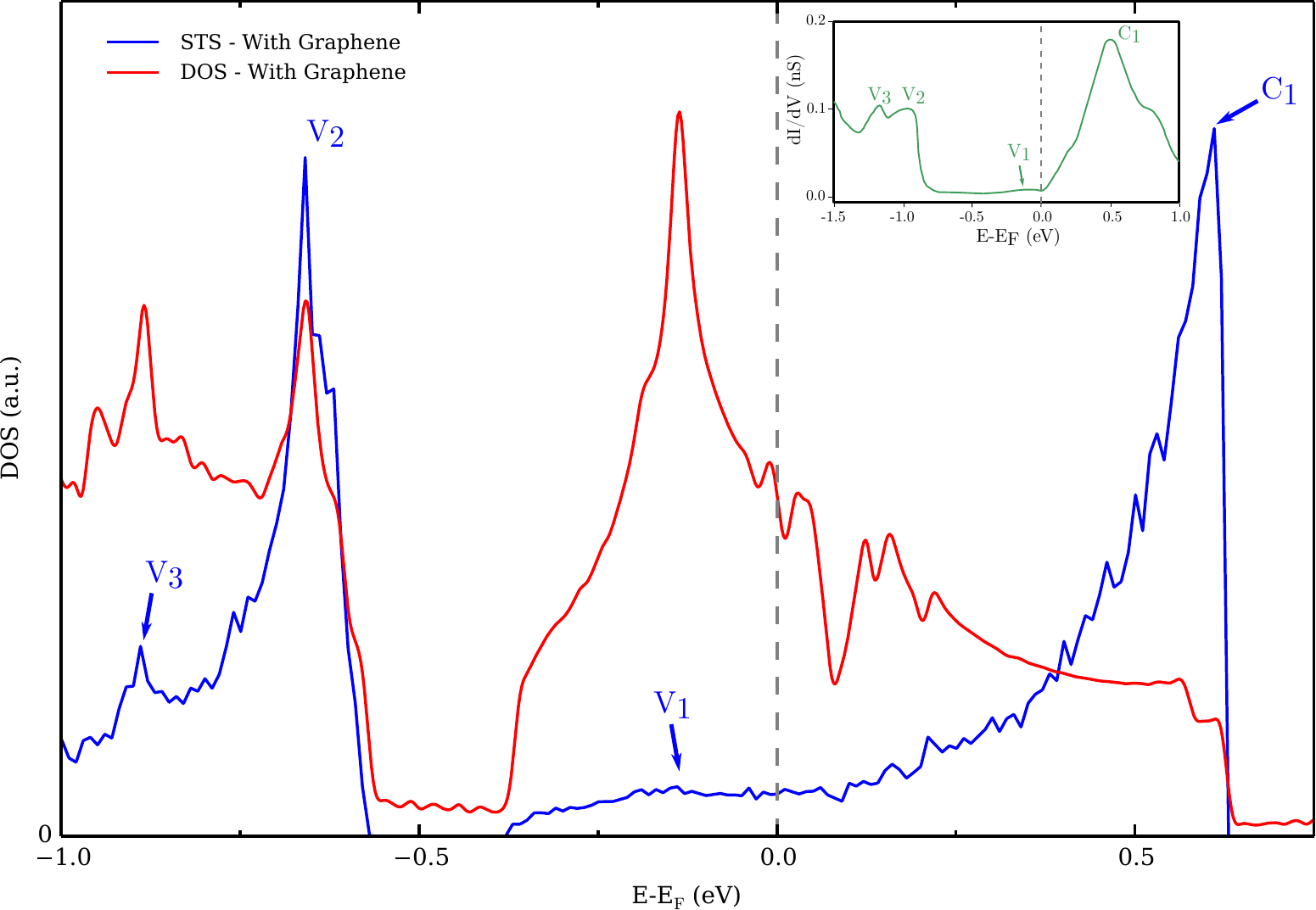}
\caption{Calculated STS spectum (blue line) and DOS (red line) for the graphene/single-layer \nbs ~system. The inset shows the experimental STS spectrum for \nbsblg~ at 5 K adapted from the work by Ugeda \emph{et al.} \cite{Ugeda_2015}.}\label{fig:STS}
\end{figure*}

Experimentally, the gross features of the electronic structure for single-layer \nbs ~were discussed by Ugeda \emph{et al.} \cite{Ugeda_2015} on the basis of STS $dI/dV$ spectra and the ARPES measurements for the \nbsblg ~system. The experimental STS spectra from Ref.~\citenum{Ugeda_2015} is presented in the inset of Fig.~\ref{fig:STS} showing four features denoted V$_1$, V$_2$, V$_3$ and C$_1$. On the basis of a comparison with the $sp$-polarized ARPES dispersion at 300 K and band structure calculations for undistorted \nbs ~in bulk and single layer,~\cite{Calandra2009} these features were assigned by Ugeda~\emph{et al.} in the following way: (i) the very shallow V$_1$ peak was assigned to dispersive bands in the $\Gamma \rightarrow M$ and $\Gamma \rightarrow K$ lines, (ii) the peak at C$_1$ was associated with the top of the Nb-based conduction band at $\Gamma $, and (iii) the two peaks V$_2$ and V$_3$ were assigned to the lower-energy valence band structure. The results of our calculations, shown in Fig. \ref{fig:DOS_Exp}, and the accompanying discussion prove that the attribution of peak V$_1$ is essentially correct while providing considerably more detailed information. In contrast, our discussion apparently suggests that the assignment of peak C$_1$ is not appropriate because of the small weight of the  Nb band around the $\Gamma $ point can not be responsible for the very large C$_1$ peak. Also, in agreement with the experimental results, we find that slightly below the region associated with peak V$_1$, a gap ($\sim $ 0.23 eV) opens because of the absence of interlayer Se-Se interactions. To analyze the origin of the V$_2$ and V$_3$ peaks we must take into account the effect of the underlying graphene.
The DOS calculated for the isolated single-layer \nbs ~(see Fig. \ref{fig:DOS_Exp}) exhibits only a peak at the top of the valence band and thus can not explain the occurrence of both V$_2$ and V$_3$ peaks.
However, our calculations including an underlying graphene layer (see Fig. \ref{fig:NbSe2_graphene} or the red line in Fig. \ref {fig:STS}) show that, because of the interlayer interaction, there are two nearby Se based peaks which we associate with the V$_2$ and V$_3$ peaks. It thus appears that, except for the region corresponding to positive bias voltages, the experimental and theoretical results are in good agreement and provide a quite detailed description of the electronic structure of  \nbs ~single-layers.
The strong maximum labeled C$_1$ is associated with a small peak of the DOS.
However, this apparent contradiction is most likely due to the neglect of the tunneling selectivity when comparing the STM $dI/dV$ spectrum with the DOS.
In order to provide a more meaningful comparison with the experimental results we have carried out an integration of the STM intensity with the energy (see the blue line in Fig. \ref {fig:STS}).
These results nicely account for all important features of the experimental STM $dI/dV$ spectrum.
A large peak occurs at the top of the conduction band (C$_1$), a weak maxima occurs in the region of the strong peak in the DOS slightly below the Fermi level (V$_1$) and two near maxima occur in the region of the top of the valence band (V$_2$ and V$_3$).
Consequently, the good agreement between the experimental and theoretical results confirm that: (i) the shallow V$_1$ peak occurring slightly below the Fermi level originates from the Nb based partially filled band in the region of the minima in the $\Gamma \rightarrow M$  and  $\Gamma \rightarrow K$ lines associated with the mixing of the Nb $z^2$ and $x^2$-$y^2$/$xy$ orbitals and leading to the stabilization of the extended Nb-Nb interactions; (ii) the strong peak around 0.5 eV is associated with the Nb $z^2$ orbitals which heavily dominate the wave functions around the $\Gamma $ point, leading to antibonding Nb-Nb and Nb-Se interactions, and (iii) the V$_2$ and V$_3$ peaks are associated with the two strongly Se $p$ based levels, the lower one arising because of the interaction with graphene.

\section{Concluding Remarks}
The electronic structure of \nbs ~single-layers in the normal non-modulated and the $3\times3$ CDW states has been studied in detail. It is shown that the CDW barely affects the Fermi surface of the system thus ruling out the much debated nesting mechanism as the driving force for the modulation. In addition, a detailed comparison of the band structures and DOS shows that the CDW stabilizes levels lying around 1.5 eV below the Fermi level that, even being in the Se $p$-block bands, have a substantial Nb-Nb bonding character. The scanning tunneling microscopy simulations carried out clearly show that the actual images noticeably change with the sign and magnitude of the voltage bias. Thus, although the corrugation of the Se sublayer induced by the modulation plays a primary role in leading to these images, the electronic reorganization has also an important contribution. The more/less brilliant spots of the STM images are associated with the Se atoms sitting on top of a triangle of Nb atoms with shorter/longer Nb-Nb contacts induced by the CDW distortion. It is also shown that the underlying graphene layers used in experimental studies of the \nbs ~single-layers, do not modify the conduction valence states but have some influence on the shape of the valence band. These results have been used to analyze recent experimental STS spectra providing a very detailed description of the electronic structure of \nbs ~single-layers.  According to a recent proposal,~\cite{Noat2015} a strong quasi-particle coupling occurs between the Se based pancake and the K-centered cylinders of the Fermi surface in the bulk leading to the larger of the two superconducting gaps of $2H$-NbSe$_2$. Consequently, the absence of the Se-based pancake component of the Fermi surface for single-layer \nbs ~as a consequence of the absence of interlayer interactions, must be related to the variation of its superconducting properties. 

\section*{Acknowledments}

This work has received funding from the European Union's Seventh Framework Programme (FP7/2007-2013) through the ERC Advanced Grant NOVGRAPHENE (GA 290846). Work in Bellaterra was supported by Spanish MINECO (Grant Nos. FIS2015-64886-C5-3-P and FIS2015-64886-C5-4-P, and  the Severo Ochoa Centers of Excellence Program under Grants SEV-2013-0295 and SEV-2015-0496), and Generalitat de Catalunya (2014SGR301).  We thank M. Ugeda for fruitful discussions.

\appendix

\section{Computational details}\label{sec:app_compdet}

DFT calculations were carried out using a numerical atomic orbitals density functional theory 
approach~\cite{HohKoh1964,KohSha1965}, which was developed for efficient calculations in large systems and implemented in the \textsc{Siesta} code~\cite{SolArt2002,ArtAng2008}.
We have used the generalized gradient approximation (GGA) to DFT and, in particular, the functional of Perdew, Burke and Ernzerhof~\cite{PBE96}.
Only the valence electrons are considered in the calculation, with the core being replaced by norm-conserving scalar relativistic pseudopotentials~\cite{tro91} factorized in the Kleinman-Bylander form~\cite{klby82}.
The non-linear core-valence exchange-correlation scheme~\cite{LFC82} was used for all elements.
We have used a split-valence double-$\zeta $ basis set including polarization functions, optimized for the bulk structure of NbSe$_2$~\cite{arsan99}.
The energy cutoff of the real space integration mesh was 300 Ry.
The Brillouin zone was sampled using a grid of (100$\times$100$\times$1) {\it k}-points within the Monkhorst-Pack scheme~\cite{MonPac76}.
Except otherwise stated our single-layer results are based on the experimental crystal structure obtained by Malliakas \etal~\cite{malliakas_2013} for the CDW structure of \nbs ~in the bulk. 

Since the bonding for the graphene/single-layer \nbs ~is supposed to be of van der Waals character and it is well known that GGA does not show any binding for this kind of systems, we used the local density approximation (LDA) to determine the interlayer distance. LDA has a tendency to overestimate the binding energy from the superposition of the charge densities of the two layers and, from previous experience, it is known that provides results not far from the experimental ones, due to the cancelation of errors between the overbinding and the lack of true van der Waals dispersion energies.
The lattice mismatch between graphene and \nbs ~forces us to build large supercells to have commensurate periodic structures. As a first approximation to simulate the graphene/single-layer \nbs ~system we built a $4\times4$-unit cell of graphene and place it on top of the $3\times3$ \nbs ~single-layer. The graphene/single-layer \nbs ~separation was optimized finding the minimum of the energy versus distance, which is found to be 3.38 \AA. Such distance was checked to be correct by doing additional calculations using a $17\times17$-unit cell for graphene.

For the STM images we use the Tersoff-Hamman approximation,~\cite{TerHam85} where the current at a given tip position is proportional to the LDOS at that point, integrated over the standard energy window given by the tip-surface potential difference ($E_F, E_F+eV$). 
Our images correspond to constant current images, showing the maps of heights that produce a constant tip-surface current.
Instead of specifying the value of the current (which is the situation encountered in an experiment), we choose a particular value of the density of states and plot the corresponding constant DOS surface.

\bibliography{dft_nourl,NbSe2_1Layer_nourl}

\end{document}